# Robust protein-protein interactions in crowded cellular environments

Eric J. Deeds[1], Orr Ashenberg[2], Jaline Gerardin[3] and Eugene I. Shakhnovich[4]

*1) Department of Systems Biology, Harvard Medical School, 200 Longwood Ave., Warren Alpert #536, Boston, MA 02115*

*2) Computational and Systems Biology Program, Massachusetts Institute of Technology, 77 Massachusetts Avenue, Building 68, Cambridge, MA 02139*

*3) Harvard College, 12 Oxford Street, Cambridge, MA 02138*

*4) Department of Chemistry and Chemical Biology, Harvard University, 12 Oxford Street, Cambridge, MA 02138*

**The capacity of proteins to interact specifically with one another underlies our conceptual understanding of how living systems function. Systems-level study of specificity in protein-protein interactions is complicated by the fact that the cellular environment is crowded and heterogeneous; interaction pairs may exist at low relative concentrations and thus be presented with many more opportunities for promiscuous interactions compared to specific interaction possibilities. Here we address these questions using a simple computational model that includes specifically designed interacting model proteins immersed in a mixture containing hundreds of different unrelated ones; all of them undergo simulated diffusion and interaction. We find that specific complexes are quite robust to interference from promiscuous interaction partners, only in the range of temperatures $T_{design} > T > T_{rand}$. At $T > T_{design}$ specific complexes become unstable, while at $T < T_{rand}$ formation of specific complexes is suppressed by promiscuous interactions. Specific interactions can form only if $T_{design} > T_{rand}$. This condition requires an energy gap between binding energy in a specific complex and set of binding energies between randomly associating proteins, providing a general physical constraint on evolutionary selection or design of specific interacting protein interfaces. This work has implications for our understanding of how the protein repertoire functions and evolves within the context of cellular systems.**

# Introduction

Protein interactions represent one of the most fundamental and important biophysical modalities within the cell. The concept that proteins have the ability to interact with both high specificity and affinity within the cell has long provided the underlying biophysical basis for much of our understanding, conceptualization and modeling of signal transduction and cellular function. The central importance of protein-protein interactions (PPIs) has lead to extensive study of this phenomenon, both experimentally (1-4) and theoretically (5-12).

The introduction of high-throughput methods for determining PPIs, such as the yeast-2-hybrid (Y2H) technique, has provided a new perspective on understanding and modeling interactions themselves. These experiments, while by their nature very noisy and potentially unreliable (1, 2, 13, 14), have nonetheless provided a remarkable first glimpse into the set of protein interactions that might exist within a living cell. These networks have also highlighted the fact that, while our understanding of very particular protein interactions might be quite advanced (8, 15, 16), our understanding of how specific interactions arise and fare dynamically within the cell is still in its infancy. Recent work has provided some insights into how large-scale protein-protein interaction networks might be dynamically organized using information from structurally characterized complexes (9-12). This work clearly indicates that the cellular environment contains a large number and variety of molecular species, many of which have the opportunity to interact with one another either specifically or non-specifically. These observations immediately beg the question of how a pair of proteins that have evolved to interact might actually behave in cytoplasm. This problem is confounded by the fact that many proteins occur at surprisingly low copy number and relative concentrations (17), which implies that such proteins will encounter promiscuous interaction partners much more frequently than the few protein molecules with which they have evolved to specifically interact. Indeed, it was recently shown that the topological features of Y2H PPI networks can be completely reproduced by a non-specific model of protein interactions based solely on hydrophobicity (14). Recently Sear studied a simple lattice 2*2*2 model of many interacting proteins where protein surfaces are presented as contiguous patches of hydrophobic or polar residues (18). He found that in this model of purely hydrophobic binding the capacity of proteins to form specific interactions is limited essentially by combinatorics of surface patches and that in turn may limit the size of the proteome. These experimental and theoretical results highlight the need to understand protein interactions in the dynamic context of the cell.

In this work we address some of the above questions in a direct biophysical fashion. To do so requires that one explore a system in which many different molecular species are considered at once. Much like early work on protein folding, we rely on a reduced representation of proteins, namely, lattice polymers (19), to overcome some of the inherent theoretical and computational difficulty involved with simulating and understanding the behavior of large ensembles of (distinct) proteins interacting with one another. In this case, hundreds of lattice proteins (hereafter referred to as simply "proteins") undergo simulated, Monte Carlo diffusion in a 3-D lattice space; interactions between different protein molecules are mediated by a standard residue-residue interaction potential for protein folding. One may populate this lattice space with a wide variety of different lattice proteins. In this case, we focus on a pair of lattice proteins that have been "designed" to interact with one another specifically. The behavior of this pair is compared to the behavior of large sets of "random" proteins that have not been similarly designed; the behavior of this specific interaction is also evaluated in the context of a background of such random proteins at varying relative concentrations. This approach, while inherently coarse-grained and approximate, nonetheless captures a number of the salient features of the problem of the formation of specific PPIs in a concentrated, heterogeneous environment. The lattice proteins in question, while not large, nonetheless present a large enough surface that specific interactions may be energetically designed, and the nature of the simulation allows one to consider interaction specificity at the ensemble level as a function of both concentration and temperature.

Study of this model reveals a number of surprising and important findings. Firstly, the fraction of proteins involved in a complex undergoes an approximately first order phase transition with temperature; interestingly, this transition occurs both for sets of designed *and* "random" proteins, although the transition temperature itself depends on the actual identities of the molecules themselves. At intermediate temperatures the specific interactions between designed pairs are maintained even at very low relative concentrations; that is, specific protein complexes can be very robust to interference from other potentially promiscuous interactions despite the fact that opportunities for such interactions are far more common than opportunities for specific interactions. At lower temperatures, however, promiscuous interactions become far more prevalent, interfering with specific interactions even at modest concentrations. These results have profound implications for our understanding of specificity in protein interactions in the cell and make specific predictions regarding optimal temperatures for the conduct of high-throughput PPI experiments.

## Results

### Interaction-Diffusion Algorithm

In order to investigate the behavior of large ensembles of interacting molecules, we developed a Monte Carlo Interaction-Diffusion (MCID) algorithm as a simplified model of ensemble protein interaction dynamics. This simulation executes a simultaneous random walk for a large set of lattice proteins in the 3-D cubic lattice. The simulated space exhibits periodic boundary conditions and can be populated by 100s to 1000s of different molecules. The proteins are treated as inelastic objects with a fixed shape: that is, excluded volume is strictly maintained in the simulation and the polymers themselves do not unfold or undergo other internal conformational transitions.

The proteins we employ are maximally compact 27-mers (i.e. 3x3x3 cubes), a model that has been extensively studied in the context of protein folding, evolution and function (19-23). This extensive body of work allows us to design 27-mer sequences that will fold quickly, reliably and stably (20, 24, 25) into a chosen native state and also allows us to maintain stable folding when designing sequences to interact (see the Methods). We employ the Mirny-Shakhnovich (MS) (26) potential both as our potential for designing sequences and as the interaction potential between proteins in the simulation, as defined below. Although we do not explicitly model protein folding and unfolding in this simulation (given the prohibitive computational cost of tracking microscopic conformational details for 100s of individual diffusing molecules), designing sequences that fold into specific lattice conformations creates polymers with specific sets of surface residues that are available for interaction in the simulation.

In order to model interactions, we must have a method for evaluating the energy of the system in any given overall conformation. To do this we apply the protein folding potential mentioned above (i.e. the MS potential) to all of the pairwise nearest-neighbor residue-residue interactions between the polymers in the system. Formally, for a system consisting of $N$ separate proteins, the energy of a particular state in the simulation is defined as:

$$E = \sum_{i=1}^{N} \sum_{j=i+1}^{N} \left( \sum_{s_i=1}^{27} \sum_{s_j=1}^{27} \epsilon(s_i, s_j) \delta_{s_i s_j} \right) \quad (1)$$

where $i$ and $j$ sum over all the proteins in the simulation, $s_i$ and $s_j$ sum over all of the sequence positions in $i$ and $j$, $\varepsilon(s_i, s_j)$ is the energy of interaction between the residue types at sequence positions and $\delta_{ij}$ is 1 if $s_i$ and $s_j$ are neighbors on the lattice and 0 otherwise. The random walk is executed by choosing a random polymer at each time point and either translating it in a random direction or rotating it by 90° in a random direction. If the resultant move causes any "overlap" between the proteins (i.e. results in any 2 proteins that are occupying one or more of the same lattice vertices), the move is immediately rejected in order to maintain excluded volume. If such an overlap does not occur, the energy of the system is evaluated after the move. If the energy of the system decreases, the move is accepted; if it increases, the probability $p_A$ of accepting the move is determined according to the Metropolis criterion:

$$p_A \sim e^{\frac{-\Delta E}{T_S}} \quad (2)$$

where $\Delta E$ is the difference in energy of the system before and after the move and $T_S$ is the Monte Carlo simulation temperature.

The fact that proteins in this simulation should be able to form stable complexes with other proteins necessitates the inclusion of complex-specific movement rules in the algorithm. In this case, one would like the simulation to allow proteins to form complexes, to allow such complexes to dissociate (according to the above Metropolis criterion for the change in energy after dissociation) as well as accurately modeling the translational and rotational diffusion of the complexes themselves. In this case, we restricted our simulation to only allow binary complexes; the movement rules for accurately modeling complexes of arbitrary size are considerably more complex and computationally costly, and as such are beyond the scope of this work. Moves that would result in ternary complexes (i.e. moves that place a protein in contact with a protein already involved in a complex) were thus always rejected in order to prevent the formation of ternary or higher-order complexes. If a single protein involved in a complex is moved, and the move is energetically accepted, then the complex dissociates—this allows for the loss of complexes with the correct probability at a given Monte Carlo temperature. If the move is rejected (i.e. if the complex does *not* dissociate), then the protein that was chosen to move is either returned to its original position/orientation (with probability 1/2) or the other protein in the complex is moved to a position consistent with the new location of the moved protein. Should movement of the entire complex result in an orientation that violates excluded volume, both proteins are returned to their original positions. These complex-specific rules result in a situation in which "rejected" moves might actually result in a change in the state of the system; in this case, the moved protein essentially "pulls" its complexed partner around form time to time during the simulation.

Complexes thus diffuse around the space in much the way that single proteins do, but with a lower diffusion constant consistent with their larger size (27).

**Specific Heterodimers**

We first apply the MCID algorithm described above to the case of a specific heterodimer in which two different polymers are designed to interact with one another. The question of *how* to design such an interaction is an interesting one. Here we employ an approach taken from the study of protein folding and use a "binding Z-score" metric to produce an interaction between two proteins that is considerably lower in energy than the interaction of those proteins with a large set of randomly chosen polymers (with sequences that are designed to fold into specific structures; see Methods). We thus obtain a pair of proteins (here denoted as proteins *A* and *B*) that have the potential to interact in a stable and specific manner. We consider this specific heterodimer in the simplest case first; that is, in the absence of any competing interactions or other proteins.

We ran simulations at a wide variety of temperatures and concentrations. We chose an intermediate concentration of 200 total proteins (100 A and 100 B) in order to allow for ample binding opportunities in the system but without moving the system into a high concentration where crowding became an issue. In terms of the general behavior of the system, we found that concentration regimes between 50-1000 proteins in this lattice space behaved essentially identically to the 200-protein case considered extensively below (data not shown). In general, the simulations reach energetic equilibrium within ~$1 \times 10^6$ steps (depending on temperature, see the Methods and Figure 1A). The relaxation to energetic equilibrium is approximately exponential across a wide variety of temperatures (see Figure 1A for a representative set of simulations). The equilibrium energy of the system at a given temperature was determined by averaging over the final $5\times10^5$ steps from a single simulation (see Methods). The dependence of the energy of the system (defined in equation (1)) on temperature in these simulations displays a number of interesting features (Figure 1B). At low temperatures, the system becomes "frozen" and many non-specific (and relatively unfavorable) interactions are observed. These non-specific complexes involve *AA* and *BB* homodimers as well as *AB* complexes with orientations other than the one that was specifically designed (see Methods). At intermediate temperatures, the system robustly reaches a minimum energy. We also observe that the energy of the system increases dramatically over a short range of temperature values, a finding that is indicative of a first-order-like phase transition in the binding state of the system. To understand the binding behavior of the system, we define the specific binding fraction (denoted *r*) as:

$$r = \frac{N_{AB}}{\frac{1}{2}(N_A + N_B)} \quad (3)$$

where $N_{AB}$ is the number of *AB* complexes in the system and $N_A$ and $N_B$ are the total number of *A* and *B* molecules in the system. The total number of *A* and *B* molecules are always equal in these simulations, and so *r* simply corresponds to the fraction of the total possible number of *AB* complexes that are actually observed in the simulation. We find that the *r* undergoes a transition similar to that found in the energetic case (Figure 1C); the interactions in the system essentially melt once the temperature approaches a regime in which dissociation becomes probable. Consistent with this observation, we find that most complexes are comparatively long lived below the melting temperature, while above it we find that interactions are transient (data not shown).

It is interesting to note that, although the interaction between *A* and *B* is much lower in energy than the interaction of *A* or *B* with random proteins, we find that the *A-A* and *B-B* homodimers are quite close in energy to the desired *A-B* complex and, as such, occur with high frequency in these simulations. These spurious homodimers provide an explanation for the fact that the maximal value of *r* we observe is ~0.5 rather than 1 (Figure 1C). This finding is not terribly surprising given that we did not specifically design against homodimers and given the fact that homodimers are statistically more likely to be stable (28, 29). This result highlights, however, exactly *how* explicit negative design must be in order to ensure correct complex formation even in binary ensembles.

**Random Lattice Polymers**

It was recently suggested that random interactions between proteins (largely driven by the hydrophobic effect) might actually underlie many of the interactions observed in PPI networks derived from high-throughput Y2H methods (14). These observations immediately raise the question of how ensembles of proteins that have not evolved to interact with one another might behave in the cell under varying conditions.

We first considered a case in which hundreds of different random polymers were simulated together; in this case, we chose to simulate sets of 200 different proteins at once. These sequences were designed to fold with low Z-scores into each of the 200 (structurally distinct) native states, but were not designed according to any consideration of interactions (see Methods). We find that, as in the case of specific heterodimers, these random ensembles relax exponentially to energetic equilibrium (data not shown) and exhibit a sharp transition in both average energy and fraction of proteins in complexes as temperature increases (see Figures 2A and data not shown); the major difference between the random case and the designed case is that the transition occurs at considerably lower temperatures. This indicates that such an ensemble will exhibit few, if any, stable interactions between random pairs at temperatures at which the interactions between heterodimers persist. Although this difference is to be expected (given that the average interaction energy among these random polymers much less favorable than the designed interaction between *A* and *B* discussed above) it is nonetheless surprising to see that the qualitative transition behavior is similar in the designed and heterogeneous random polymer cases.

The non-specific models of protein interaction mentioned earlier correctly predict that many of the interactions between proteins in such systems are not very stable (14); that is, the network is essentially very sparse and most random pairs of proteins will not interact strongly. It is thus clear that, even should a pair of random proteins in the heterogeneous ensemble exhibit a strong interaction energy, each and every such pair will be found only once and thus face an entropic barrier to forming a complex. To investigate the behavior of pairs of random proteins at high relative concentrations (similar to the situation explored for the specific *A*/*B* case), we chose two different pairs from the above ensemble at random and simulated the system with 100 of each. In these cases, we observed the binding transition to occur at temperatures intermediate between the transition for the heterogeneous ensemble and the *A*/*B* pair (Figure 2B). In this case it is clear that many random pairs of proteins, especially when they occur at appreciable concentrations (such as the overexpression employed in Y2H experiments), will form stable complexes across a variety of temperatures. This finding is consistent with the hypothesis that many interactions within Y2H datasets might consist of this type of non-designed random interaction between proteins (14).

**Competition**

The lattice interaction framework allows us to ask whether or not a designed interaction such as the *A-B* heterodimer will interact reliably in the presence of competing random interaction partners. To approach this question we ran simulations at varying relative *A/B* concentrations, starting from a system of 200 *A/B* proteins and replacing 2 *A* and 2 *B* proteins with randomly chosen proteins to obtain a range of relative concentrations from 100% to 0%. In this case, each random protein represented a unique sequence-structure pair, resulting in a completely heterogeneous random ensemble. One hundred simulations were run at each concentration at 3 different temperatures (see Methods for further details). The results of these calculations are summarized in Figure 3.

The behavior of this competitive system depends quite strongly on temperature. At the lowest temperature we considered, the random proteins have a considerable effect, causing a roughly linear decrease in the % of designed *A/B* protein pairs involved in the specific *AB* complex as the relative concentration of random competing partners increases (Figure 3A). The increasing concentration of random polymers represents an entropic contribution; both *A* and *B* are much more likely to encounter random polymers than their designed interaction partners when their relative concentration becomes small. At this temperature, the interaction between the specific polymers and members of the random ensemble are stable enough that this entropic effect is able to overwhelm the considerably greater stability of the *AB* complex.

At intermediate temperatures (Figures 3B and C), the situation changes dramatically. The designed polymers tend to participate in specific complexes rather than interact with random proteins across a wide range of relative concentrations. Indeed, at the highest temperature we considered, most *A/B* molecules were found in the specific complex even when only 8 or 6 of those molecules were included in the simulations (see Figures 3B and C at relative concentrations of the random proteins of around 90%). These findings indicate that stable, evolved interactions are remarkably robust to the interference of random polymers as long as the temperature is sufficiently high.

The above case represents the competition between a specific interaction and a completely heterogeneous ensemble where (at low relative concentrations of *A* and *B*) each individual sequence type is rare. Some proteins, however, can be expressed at very large concentrations, especially in certain overexpression experiments based on engineered constructs (1, 2). To explore the effects of a high relative concentration of a *single* polymer on specific interactions, we chose a single sequence-structure pair from the non-interacting set and performed simulations analogous to those shown in Figure 3 (see Methods). The single polymer was taken arbitrarily from the non-interacting set without consideration of its average hydrophobicity or any other characteristic that might influence its ability to bind non-specifically. The results of the simulations at three different temperatures are shown in Figure 4.

As can be seen from comparing Figures 3 and 4, the *AB* interaction is far more susceptible to competition from a single random polymer than from a completely heterogeneous ensemble at every temperature. At low and intermediate temperatures, the *AB* interaction becomes unlikely (>10%) even when the relative concentration of *A* and *B* is around 20%, while at high temperatures the specific complex demonstrates robustness comparable to that seen at intermediate temperatures with the heterogeneous ensemble. From these results it is clear that robust interaction specificity will only manifest itself in a narrow temperature range—namely, the range in which the average interaction of the designed proteins is unstable relative to the temperature but the designed interaction is stable. High relative concentrations of any given protein (especially proteins that have high surface hydrophobicity or are overly "sticky") may narrow this temperature range considerably, a fact which may explain some of the unexpected or non-specific effects of overexpression.

**Discussion**

Although the simulations described above represent only a first approximation to the statistical dynamics of protein interactions in the crowded environment of the cell, they nonetheless provide a number of interesting insights into how such a system might behave. The lattice polymers and diffusion-interaction system studied here captures a number of the fundamental features of the problem of interaction specificity in heterogeneous environments. These polymers exhibit realistic surface residue statistics and represent proteins that can actually fold (at least within the context of a lattice model). The interaction potential employed is the same as that used to design the sequences for folding and represents a plausible representation of interaction energetics; this model accounts explicitly for many features of residue-residue interaction energies and implicitly models the influence of solvent and the hydrophobic effect (20, 26). The MCID algorithm thus allows us to obtain a first explicit picture of how evolved interactions might actually fare when the components of such complexes are at low relative copy number and are presented with numerous opportunities for nonspecific interactions.

The key finding of this work is that certain degree of evolutionary design of specific protein-protein interaction strength is necessary to avoid loss of specific protein complexes to possible promiscuous interactions with abundant other proteins in the cell. The physical manifestation of this phenomenon is in the separation of two transition temperatures: temperature of co-condensation transition of random protein surfaces ($T_{rand}$) (predicted in (30)) and temperature at which specific complexes lose stability ($T_{specific}$). Our simulations and analysis show that specific interactions can form only in a limited temperature range

$$T_{specific} > T > T_{random}$$

This temperature window may exist only when specific surfaces are designed for strong recognition. For example, in a ''one-shot'' selection scenario where interacting proteins are drawn from a primordial ''random soup'' of exposed protein surfaces $T_{specific} = T_{rand}$. Apparently design of interacting surfaces with large Z-scores as presented here creates a large energy gap between specific interaction and random association providing a finite temperature range at which specific associations occur with high fidelity and without interference from association with random protein surfaces. A conceptually similar situation is found in protein folding where native conformation should be thermodynamically stable despite *a* huge number of misfolded conformations. As it is the case here, this problem in protein folding is solved by selection of sequences that provide large (extensive in protein's length) energy gap between the native conformation and the set of misfolded conformations(31-33).

The temperature dependence of specificity of protein-protein interactions provides further insights into the constraints operating on the evolution and design of interacting proteins. If temperatures are too low (or, conversely, if many proteins in the cell are relatively "sticky" in a given environmental condition), our results indicate that even the most stable and specific interactions will have significant difficulty forming reliably, especially at lower relative concentrations. Organisms living at a particular temperature must either evolve their proteome to prevent such interference (i.e. by decreasing the overall "stickiness" of commonly expressed proteins, therefore lowering $T_{rand}$) or express those proteins that must interact specifically at higher concentrations. Given the fact that protein concentration can be subject to rather strong fluctuations (17, 34), in these situations cells run the risk of randomly moving between concentration regimes at which complexes exist to one in which they do not tend to form at all. However, in cases where such evolutionary constraints have resulted in a proteome that is less "sticky" in the temperature range experienced by the cell, energetic design for specificity results in a remarkable robustness in complex formation to concentration variations. Cells in this regime will exhibit specific complexes regardless of fluctuations in copy number and will have the freedom to tailor expression levels according to other evolutionary concerns.

Our results have interesting implications for the experimental study and determination of protein-protein interaction networks within the cell. Methods like Y2H which rely on overexpression (1, 2) will tend to be prone to reporting nonspecific interactions; at such high relative concentrations, such interactions will occur even though they are not necessarily stable enough to occur at biologically relevant protein levels. Methods that do not involve overexpression in this manner, such as TAP-tagging (3), will not report such interactions since they are simply much more unlikely to occur. Our results also indicate that methods which allow for explicit temperature control, such as *in vivo* experiments conducted at the highest permissive temperature or the more quantitative chip-based assays (4), might allow one to move into a regime where non-specific interactions are not stable compared to background interactions but evolved interactions are.

The results presented above represent only a brief first look into the lives of protein interactions as they actually might occur in the cell. The MCID framework is a coarse-grained but powerful approach that can be readily expanded to include larger numbers and types of molecular species as well as larger complexes and structures. This framework may be applied to answer key theoretical questions regarding topological dynamics and the relationship between static information about network structure and the actual state of the system within a cell.

## Methods

**Folding Sequence Design**

The 27-mer polymers used in this study were all designed to fold into particular 3x3x3 lattice structures. For each sequence-structure pair, a conformation was chosen at random from the 103,346 different maximally compact 27-mer structures on the cubic lattice (19) to be the native state. For each chosen native state, a sequence was designed to fold into that conformation using a Z-score minimization algorithm. The folding Z-score (20, 24, 26, 31)(here called an F-score and denoted F) of a sequence i that folds to native structure j is defined in the following way:

$$F_{ij} = \frac{E_{ij} - \langle E_i \rangle}{\sigma_E} \quad (4)$$

where $E_{ij}$ is the energy of sequence i in structure j, $\langle E_i \rangle$ is the average energy of the sequence i in all the possible maximally compact 27-mer conformations and $\sigma_E$ is the standard deviation in energy across all possible conformations. The energy of sequence i in a structure j is defined as:

$$E_{ij} = \sum_{k=1}^{27} \sum_{l=k+1}^{27} \epsilon_{kl} \delta_{kl} \quad (5)$$

where k and l sum over the positions in the sequence i, $\varepsilon_{kl}$ is the potential energy of interaction between the residue types at positions k and l and $\delta_{kl}$ is 1 if positions k and l are neighbors in structure j but not neighbors in the sequence and 0 otherwise. Here the MS residue-residue interaction potential is used to define $\varepsilon_{kl}$; this potential is very similar to Miyazawa-Jernigan potential (26). The F-score represents a very well tested and robust measure of both folding kinetics and thermodynamics; sequences with low F-scores tend to fold quickly and stably into their designated native states (20, 24, 26). The design procedure began with a completely random 27-mer sequence and a Monte Carlo F-score minimization was performed on these starting sequences as described elsewhere (22). All of the sequences we designed exhibited F-scores of -7 or less, ensuring very fast and stable folding (22). Using this algorithm, we designed 3200 independent sequence structure-pairs, none of which corresponded to identical native states (i.e. all the polymers in this case were structurally unique). The random sets of polymers employed in the MCID simulations were drawn from these 3200 polymers, as were the initial sequence-structure pairs that served as the starting point for the design of the specific *A/B* heterodimer.

**Monte Carlo Interaction-Diffusion Algorithm**

Here, we explicitly considered a cubic lattice space consisting of 100 lattice vertices to a side; this 100x100x100 cube was constructed with periodic boundary conditions in order to ensure conservation of mass and of protein concentration. The move set for the algorithm is described in the main text and allows for the translational and rotational diffusion of lattice polymers in this space. All moves were constructed so that the residues of the polymers could only occupy lattice sites; rotational moves always represent 90° rotations and translational moves involve the movement of a protein a distance of one lattice unit. The direction of a rotation or a translation was chosen from the set of all possibilities with equal probability, and rotational moves were attempted with the same frequency as translational moves. These relative probabilities could easily be scaled to change the relative rates of rotational and translational diffusion.

Equation (1) in the text defines the energy of the entire simulation; it is important to note that, although the residues within the polymers themselves are interacting according to a potential as defined above, this energy is not included in the overall simulation total. We employ the same potential to evaluate protein-protein interaction energies that we use to design sequences for folding; this represents a (reasonable) hypothesis that the physics driving protein folding at the residue-residue level provide an approximation to the physics of protein-protein interaction. Although this hypothesis could be easily relaxed by employing a different potential for interactions, the lack of folding dynamics in the simulation most likely minimizes any effect that the difference between folding and association potentials might introduce into the system.

**Designing Specific Interactions**

In order to understand the behavior of specific heterodimers in the MCID framework, we required a method that would allow us to design a pair of sequences that would represent a specific interaction. To do this, we defined a binding Z-score (or B-score) analogous to the F-score described above:

$$B_{ij} = \frac{E_{ij} - \langle E_i \rangle_D}{\sigma_E} \tag{6}$$

where $E_{ij}$ is the energy of interaction between proteins $i$ and $j$, $\langle E_i \rangle_D$ is the average interaction energy of the protein $i$ with a set of proteins in a decoy ensemble $D$ and $\sigma_E$ is the standard deviation in interaction energy for the protein $i$ across that ensemble. It is important to note that the B-score is, in fact, defined only for a particular pair of surfaces from $i$ and $j$ that are interacting in a particular orientation. The term $E_{ij}$ represents the interaction energy of the two proteins in that conformation; in a sense, it corresponds to the energy $E$ obtained from equation (1) in the main text where $i$ and $j$ are the only polymers in the system and are interacting in the orientation that the algorithm would like to design into a specific interaction. The "decoy set" of energies was defined as the minimum interaction energy (across all possible interaction orientations) of the protein $i$ with each of the 3200 sequence-structure pairs designed for folding alone. The set of interaction energies is thus different for the $i$ protein compared to the $j$ protein, so, in general, we have that $B_{ij} \neq B_{ji}$, although in practice we found that the B-scores for both proteins were generally within 1 Z-score unit (data not shown).

To design a specific heterodimer, 2 proteins were chosen from the 3200 folded polymers and the interaction face was chosen to be a complete overlap between 2 of the protein surfaces (i.e. a 9x9 interaction). A Monte Carlo design procedure was employed to minimize the average B-score of the two interacting partners (i.e. $<B> = 1/2(B_{ij} + B_{ji})$). At each step, a random mutation was made at any sequence position of one of the interacting polymers. To ensure that both polymers could still fold, any mutation that brought the F-score of either polymer above -6 was immediately rejected. Those mutations that did not disrupt folding were automatically accepted if they decreased the value of $<B>$; those that increased the average B-score were accepted according to the Metropolis criterion with certain Monte Carlo design temperature $T_D$. A simulated annealing procedure was employed during B-score minimization, and the final pair of designed sequences exhibited a B-score ~ -7. These two sequence structure pairs were denoted as the *A* and *B* proteins and were employed as the specific interaction pair in this work.

**Interaction Simulations**

All of the simulations mentioned in the work were conducted for a total of $5 \times 10^6$ steps. Systems were prepared in initial conditions in which each protein to be simulated was placed in a random location and orientation within the 100x100x100 lattice space. A protein placement that violated excluded volume during the construction of these initial states was moved to a new random position until one was found that did not cause protein-protein overlap. All the initial conditions for each simulation represented independently constructed random states.

For melting simulations (i.e. those simulations leading to Figures 1B and C and Figures 2A, B and C), each point corresponds to averaging over the last $5\times10^5$ steps of 10 independent simulations at each temperature; temperature was increased from 0.5 to 10 in increments of 0.5 Monte Carlo temperature units. In each case the simulations were checked (heuristically) to ensure that energetic equilibrium had been reached before the last $5\times10^5$ steps, and at each temperature that we considered this was indeed the case.

The heterogeneous competition simulations were conducted by adding more and more proteins taken from the random ensemble of 3200 folding polymers to the *A*/*B* system. For each value of the relative concentration and temperature, we conducted 100 independent simulations and calculated the value of *r* (equation (3)) by averaging over the last $5\times10^5$ steps of these 100 simulations. In this case, each instance of the simulation contained exactly the same set of polymers; that is, the 100 simulations at any given concentration were all exactly identical in terms of their protein composition. The same relative concentrations from each separate temperature condition also correspond to the same polymer composition. Although it would have been possible to prepare the system with different sets of random competing proteins for each instance at each temperature, we chose to collect statistics on each given system rather than explore the average behavior of compositionally distinct systems. The homogeneous competition simulations involved a similar protocol, but each concentration point represents replacing a given number of copies of the *A*/*B* polymers with copies of a *single* competing protein drawn arbitrarily (and without consideration of potential interaction properties) from the heterogeneous set.

**Acknowledgements**

This work was supported by a Howard Hughes Medical Institute predoctoral fellowship to E.J.D. and the National Institutes of Health.

**Figure Legends**

**Figure 1.** Time and temperature dependence of the *A/B* ensemble. **A** System energy (defined in equation (1) in the text) as a function of Monte Carlo step in the MCID algorithm. Each curve corresponds to a different temperature. We find that the system reaches apparent energetic equilibrium within ~$10^6$ steps even at the lowest temperature considered in this work (T = 0.1). At each temperature, the relaxation to equilibrium is approximately exponential. **B** The dependence of system energy on temperature shows an initial decrease in energy (corresponding to the melting of unfavorable complexes) followed by a region of low energy (corresponding to largely *AB* complexes, although with some *AA* and *BB* homodimers) and finally undergoing a sharp transition (corresponding to the melting of the *AB* complex). **C** The dependence of *r* (equation (2) in the text) on temperature mimics that of the energy dependence in **B**. At low temperatures, many complexes are not specific, while at intermediate temperatures the system exhibits a greater fraction of specific interactions. The fact that *r* has a maximum ~0.5 is a consequence of the fact that *AA* and *BB* heterodimers are fairly stable at these temperatures, as mentioned in the text. The melting of the *AB* complex at high temperatures causes a sharp decrease in *r* as T increases.

**Figure 2.** Temperature dependence of random protein ensembles. **A** The melting transitions for 2 different heterogeneous random ensembles of proteins. In this case, the *r* term is not used to track the behavior of the system since none of the interactions are specific; the number of proteins that are in complexes overall is used as a measure of binding instead. This number decreases sharply with increasing temperature in this case, a behavior similar to that observed in Figure 1 but with a much lower transition temperature. **B** A figure similar to that in **A** but using systems composed of 100 copies each of two randomly chosen proteins (rather than a fully heterogeneous set). The two curves correspond to different choices of these 2 random potential interaction partners. The transition temperatures in this case are intermediate between the case in Figure 1 and that in **A**.

**Figure 3** Competition between a heterogeneous random ensemble and designed interactions. **A** These results are taken from simulations run at T = 0.5, a temperature at which some of the random interactions are stable (see Figure 2) but at which the energy of a pure *A/B* system is able to reach its minimum value (see Figure 1B). The black circles correspond to the measure *r* defined in the text; this is the fraction of *A/B* proteins in the specific *AB* complex. The red squares represent the fraction of all proteins in the system that are in a complex of any type, and the blue diamonds represent the fraction of complexes in the system that contain a protein from the heterogeneous random set. Notice the linear decrease in *r* (i.e. the black curve) as the relative concentration of *AB* goes from 1 (corresponding 0 on the x axis) to 0 (which corresponds to 1 on the x axis). **B** A figure similar to that in **A** but with simulations run at T = 1, which is after the melting transition in the heterogeneous ensemble but before that of the pure *A/B* system.

Note the robustness of specific interactions in this case compared to the linear decrease in the black curve in **A**.  **C**  A figure similar to that in **A** and **B** but at an even higher temperature of 1.5.  The system in this case displays even greater robustness in interaction specificity than that in **B**.

**Figure 4**  Competition between a single polymer and designed interactions.  **A**  These results were obtained from simulations at T = 0.5 as in Figure 3A. The black circles correspond to the measure *r* defined in the text; this is the fraction of *A*/*B* proteins in the specific *AB* complex.  The red squares represent the fraction of all proteins in the system that are in a complex of any type, and the blue diamonds represent the fraction of complexes in the system that contain a protein from the heterogeneous random set.  Note that the specific interaction is highly susceptible to competition from the random interaction in this case.  **B**  A figure similar to that in **A** but at T = 1.0.  At this temperature, the specific ensemble shows a linear decrease in prevalence with increasing concentration of the random interacting partner rather than the robustness observed with the heterogeneous ensemble in Figure 3B.  **C**  A figure similar to that in **A** but at T = 1.5.  In this case, the specific interaction is robust to decreases in relative concentration as in Figure 3C.

**Figure 1**

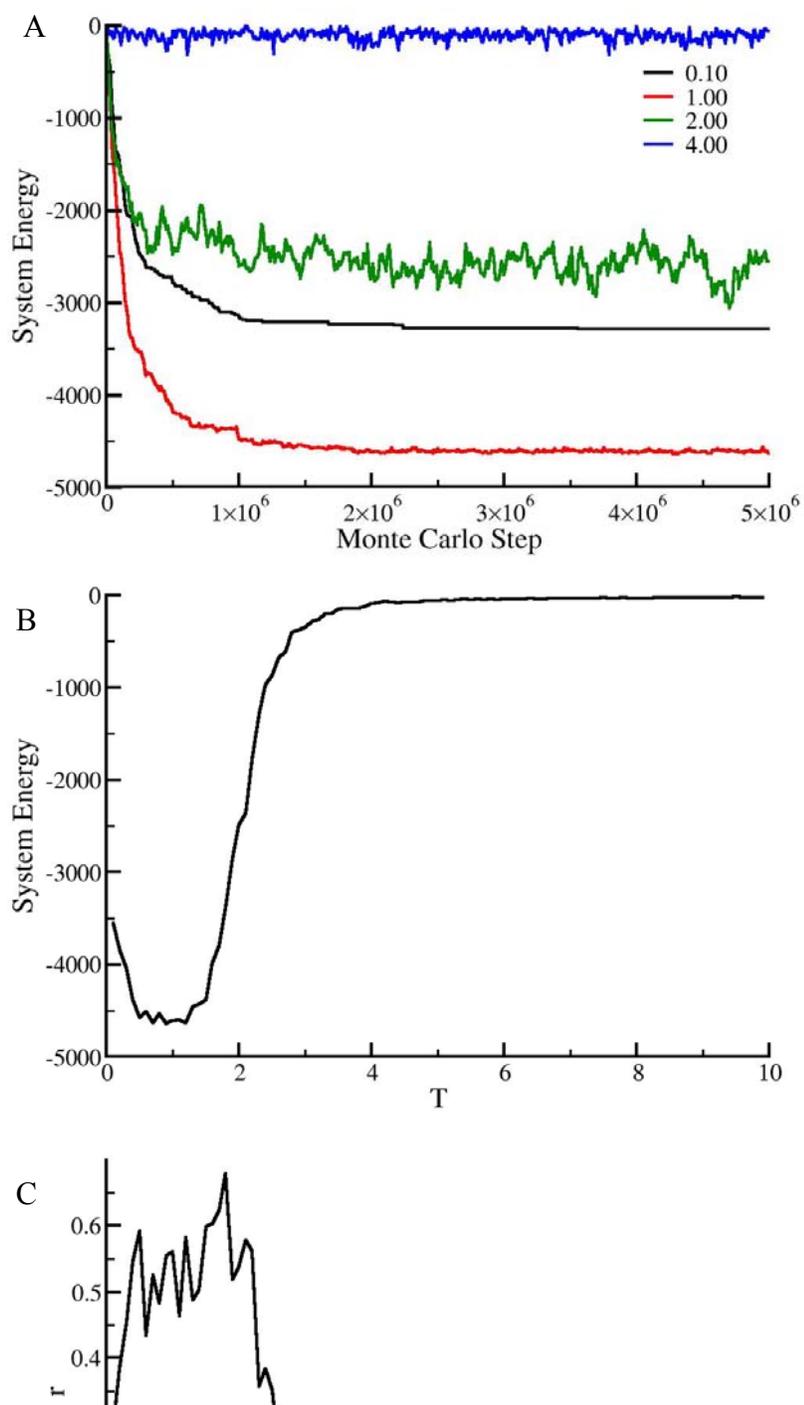

**Figure 2**

A
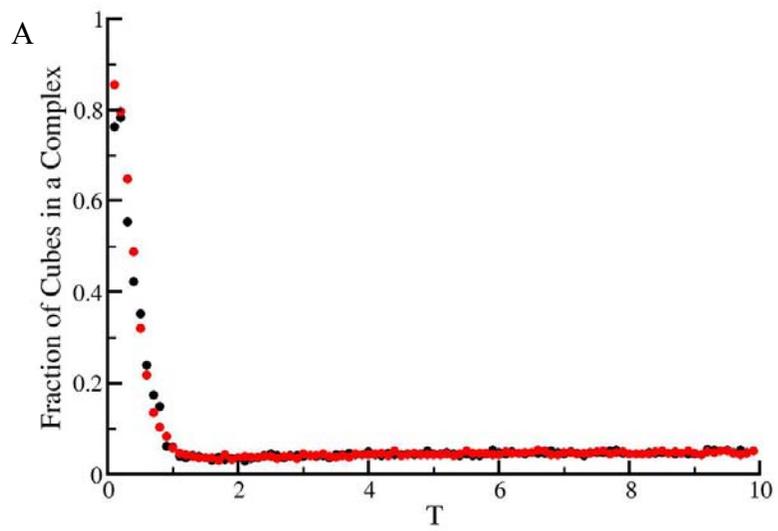

B
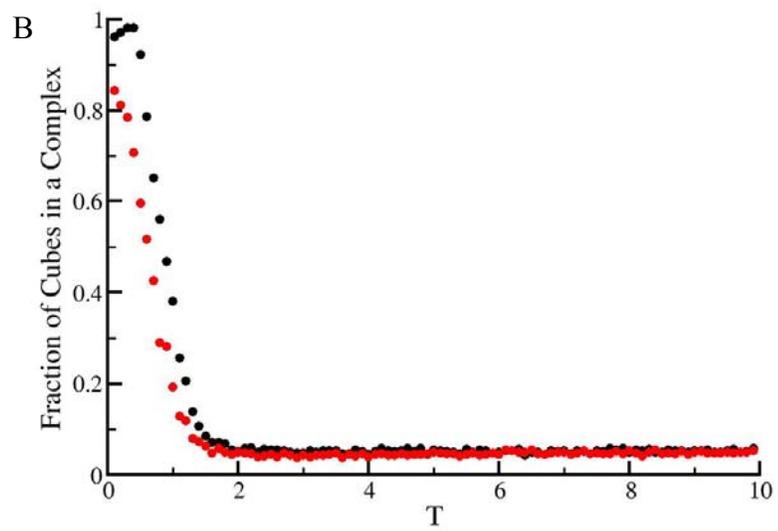

**Figure 3**

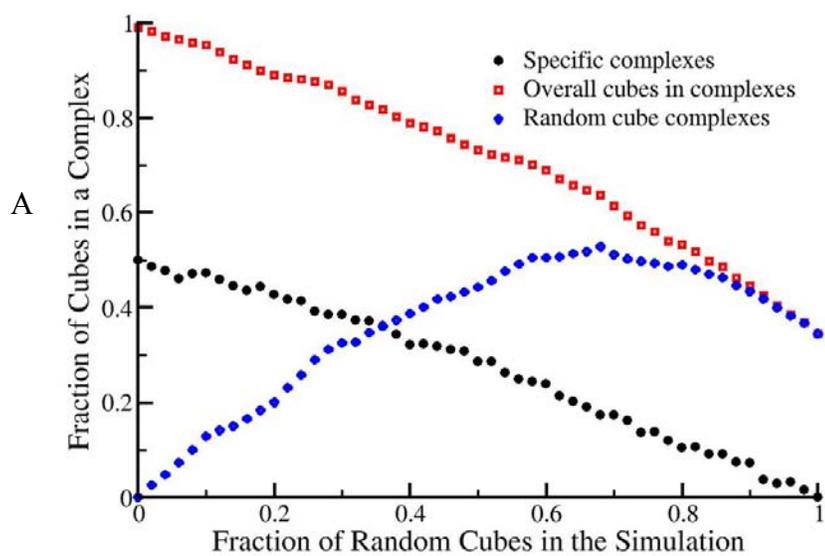

A

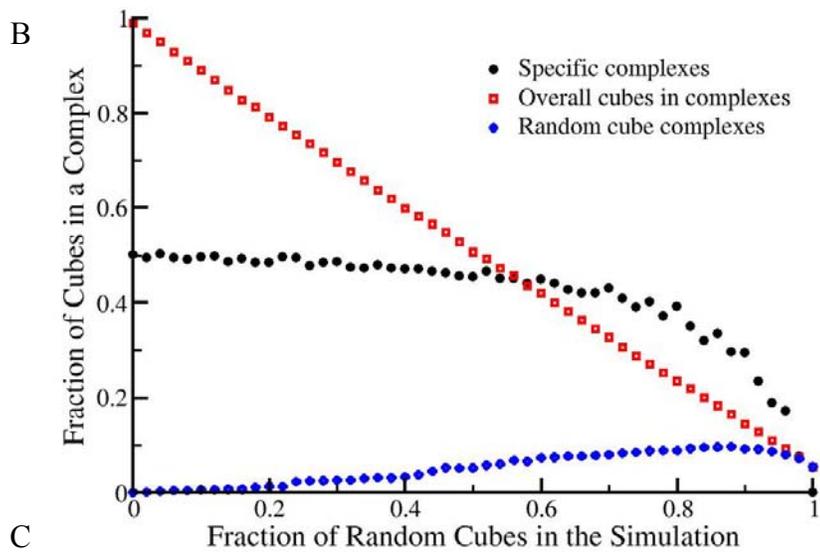

B

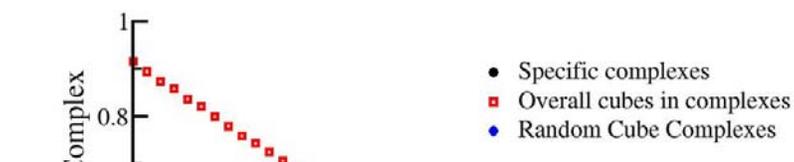

C

**Figure 4**

A
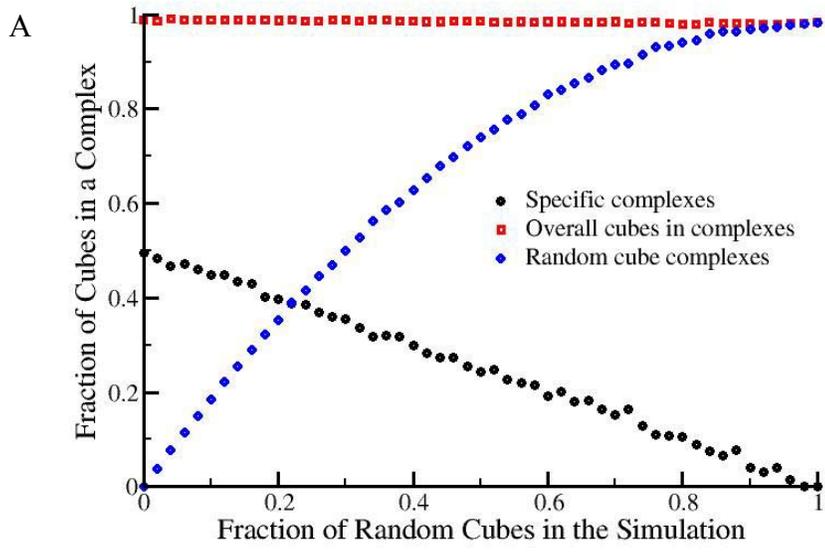

B
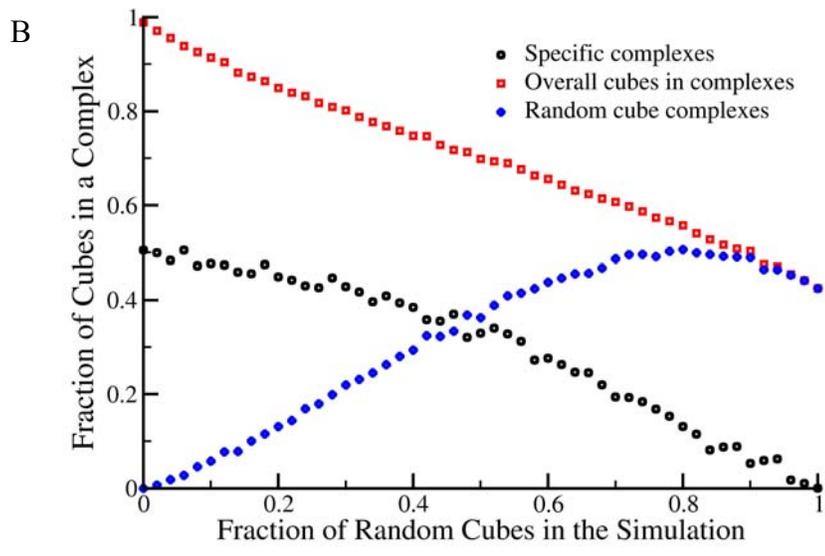

C
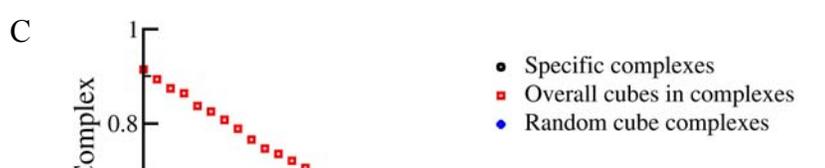